\documentclass{ws-ijmpd}
\usepackage[super,compress]{cite}
\usepackage{hyperref}
\hypersetup{colorlinks,urlcolor=black,citecolor=black,linkcolor=black,filecolor=black}
\usepackage{breakurl}
\usepackage{graphicx}

\begin{document}

%
\catchline{}{}{}{}{}
%

\title{EXPLORING THE STRING AXIVERSE AND PARITY VIOLATION IN GRAVITY WITH GRAVITATIONAL WAVES}

\author{DAISKE YOSHIDA}

\address{Physics Department, Kobe University, \\
Kobe, 657-8501,
Japan\footnote{dice-k.yoshida@stu.kobe-u.ac.jp}}

\author{JIRO SODA}

\address{Physics Department, Kobe University, \\
Kobe, 657-8501,
Japan\footnote{jiro@phys.sci.kobe-u.ac.jp}}

\maketitle

\begin{history}
\received{Day Month Year}
\revised{Day Month Year}
\end{history}

\begin{abstract}
We show that the parametric resonance of gravitational waves occurs
due to the axion coherent oscillation and the circular polarization
of gravitational waves is induced by the Chern-Simons coupling. However,
we have never observed these signatures in the data of gravitational
waves. Using this fact, we give stringent constraints on the Chern-Simons
coupling constant $\ell$ and the abundance of the light string axion.
In particular, we improved the current bound $\ell\leq10^{8}\,\,{\rm km}$
by many orders of magnitude. 
\end{abstract}


\ccode{PACS numbers: 04.30.Db, 04.30.Nk, 04.30.Tv, 04.50.Kd, 04.80.Cc, 95.35.+d}


\section{Introduction}

The direct detection of the gravitational waves (GWs), GW150914, has
opened a new window for observing the universe \cite{Abbott:2016blz}.
Thus, we can utilize GWs for exploring not only astronomy but also
fundamental physics. One of important issues in fundamental physics
is to establish quantum theory of gravity. As is well known, string
theory is a promising candidate of quantum theory of gravity. Hence,
it would be worth studying GWs to extract a hint for profound understanding
of string theory. In this letter, we will focus on the string axiverse
\cite{Arvanitaki:2009fg} which is one of fundamental aspects of string
theory.

Remarkably, the mass of string axions can take the values in the broad
range from $10^{-33}\,\,{\rm eV}$ to $10^{-10}\,\,{\rm eV}$ \cite{Svrcek:2006yi,Arvanitaki:2009fg}.
In order to explore this string axiverse, we can use the cosmic microwave
background radiations (CMB) \cite{Hlozek:2017zzf}, the large scale
structure of the universe \cite{Marsh:2010wq}, pulsar timing arrays
\cite{Khmelnitsky:2013lxt,Porayko:2014rfa,Aoki:2016mtn}, interferometer
detectors \cite{Aoki:2016kwl,Aoki:2017ehb}, the dynamics of binary
system \cite{Blas:2016ddr}, black hole physics \cite{Arvanitaki:2010sy,Yoshino:2015nsa}
and nuclear spin precession \cite{Abel:2017rtm}. The present work
intends to add an item to this list.

One of the important aspects of string axiverse is that string axions
can be the dark component of the universe. Indeed, recently, the possibility
that the axion can replace the cold dark matter (CDM) has been intensively
studied \cite{Marsh:2015xka,Hui:2016ltb}. As is well known, supersymmetric
particles, the so-called neutralinos, have been regarded as the most
promising candidate for the CDM. However, there was no signature of
supersymmetry at the LHC. Moreover, although the CDM works quite well
on large scales, there exist problems on small scales. Hence, it is
worth investigating the axion dark matter instead of the CDM. In particular,
the axion with the ultralight mass $10^{-22}\,\,{\rm eV}$ can resolve
the small scale problem of the cold dark matter \cite{Hu:2000ke}(
see \cite{Lee:2017qve} for earlier history). In fact, the axion with
any mass can behave as the cold dark matter on large scales because
of the coherent oscillation.

In the presence of the axion dark matter, it is natural to consider
the Chern-Simons terms both in the gauge sector and in the gravity
sector \cite{Campbell:1990fu,Lue:1998mq}. In particular, the gravitational
Chern-Simons term induces a coupling between the GWs and the axion
\cite{Jackiw:2003pm,Alexander:2009tp}. It should be noted that the
Chern-Simons coupling provides a natural mechanism for parity violation
in gravity provided nontrivial profile of the axion field. In fact,
the parity violation in gravity is also phenomenologically intriguing
\cite{Satoh:2007gn,Contaldi:2008yz,Takahashi:2009wc}. Another key
ingredient of the axion dark matter is the coherent oscillation of
the axion field. Since the axion is coherently oscillating, the occurrence
of parametric resonance due to the Chern-Simons coupling can be expected\cite{Soda:2017sce}.

In this letter, we examine this possibility and find a new way to
explore the string axiverse. It should be noted that the detectable
frequency range of GWs with ground based interferometers is from $1\,\,{\rm Hz}$
to $10^{4}\,\,{\rm Hz}$, which corresponds to the axion mass range
from $10^{-14}\,\,{\rm eV}$ to $10^{-10}\,\,{\rm eV}$. The GWs in
this relevant frequency range can undergo the coherent oscillation
of the axion in the center of the Galaxy where the density of the
dark matter is about $10^{3}$ times higher than the edge of the galaxy
where we live. During the journey, the GWs can be enhanced by the
resonance. Since the resonance process violates parity, there should
be parity-violation in GWs. Thus, if the chiral gravitational waves
are observed \cite{Seto:2007tn}, that indicates the existence of
the axion dark matter. When we do not observe the chiral gravitational
waves, we can constrain the Chern-Simons coupling constant or the
abundance of the string axion in the relevant mass range. Hence, it
is worth investigating the mechanism in detail.

\section{Model}

The model we consider is the dynamical Chern-Simons gravity coupled
with the axion \cite{Jackiw:2003pm}. We adopt the natural unit $c=\hbar=1$
and the conventions in \cite{Alexander:2009tp}. We choose the coordinate
$x^{\mu}\equiv(x^{0},x^{i})=(\eta,x^{i})$ with the conformal time
$\eta$ and $x^{i}=(x^{1},x^{2},x^{3})=(x,y,z)$. Then, the action
is given by 
\begin{equation}
S=S_{{\rm EH}}+S_{{\rm CS}}+S_{\Phi}\ ,
\end{equation}
where the Einstein-Hilbert action reads 
\begin{equation}
S_{{\rm EH}}\equiv\kappa\int_{\mathcal{V}}dx^{4}\sqrt{-g}R.
\end{equation}
Here, $\kappa$ is proportional to the inverse of Newton constant,
$\kappa\equiv(16\pi G)^{-1}$ and $R$ is a Ricci scalar. The Chern-Simons
term $S_{{\rm CS}}$ is given by 
\begin{equation}
S_{{\rm CS}}=\frac{1}{4}\alpha\,\int_{\mathcal{V}}dx^{4}\sqrt{-g}\,\Phi\tilde{R}R\ ,
\end{equation}
where $\Phi$ is the axion field, $\alpha$ is a coupling constant,
and $\tilde{R}R$ is the Pontryagin density defined as 
\begin{equation}
\tilde{R}R\equiv\tilde{R}_{\,\,\beta}^{\alpha\,\,\,\gamma\delta}R_{\,\,\,\alpha\gamma\delta}^{\beta}\hspace{1em}\text{and}\hspace{1em}\tilde{R}_{\,\,\beta}^{\alpha\,\,\,\gamma\delta}\equiv\frac{1}{2}\epsilon^{\gamma\delta\rho\sigma}R_{\,\,\,\beta\rho\sigma}^{\alpha}.
\end{equation}
Note that, since the axion field $\Phi$ have the dimension of the
mass, the Chern-Simons coupling constant $\alpha$ have the dimension
of length. Hence, it is convenient to express the coupling constant
as 
\begin{equation}
\alpha=\sqrt{\frac{\kappa}{2}}\ell^{2}\ ,
\end{equation}
where $\ell$ has the dimension of the length \cite{Okounkova:2017yby}.
The current limit coming from the Solar System test is $\ell\leq10^{8}\,\,{\rm km}$
\cite{AliHaimoud:2011fw}. The future experiment might improve the
contraint on $\ell$ by sixth order stronger than the Solar System
constraint \cite{Yagi:2012vf}. The action $S_{\Phi}$ of the axion
field $\Phi$ is given by 
\begin{equation}
S_{\Phi}\equiv-\frac{1}{2}\int_{\mathcal{V}}dx^{4}\sqrt{-g}\left[g^{\mu\nu}\left(\nabla_{\mu}\Phi\right)\left(\nabla_{\nu}\Phi\right)+2V(\Phi)\right]
\end{equation}
where $V(\Phi)$ is the potential of the axion field.

From the action, we obtain the equations for the metric field 
\begin{equation}
G_{\mu\nu}+\frac{\alpha}{\kappa}C_{\mu\nu}=\frac{1}{2\kappa}T_{\mu\nu}\ ,
\end{equation}
where $G_{\mu\nu}$ is the Einstein tensor and $C_{\mu\nu}$ is defined
as 
\begin{equation}
C^{\mu\nu}\equiv(\nabla_{\!\alpha}\Phi)\epsilon^{\alpha\beta\gamma(\mu}\nabla_{\gamma}R_{\,\,\,\,\beta}^{\nu)}+(\nabla_{\!\alpha}\nabla_{\!\beta}\Phi)\tilde{R}^{\beta(\mu\nu)\alpha}\ .
\end{equation}
The trace of this tensor identically vanishes. The energy momentum
tensor $T_{\mu\nu}$ becomes 
\begin{equation}
T_{\mu\nu}=\left[(\nabla_{\mu}\Phi)(\nabla_{\nu}\Phi)-\frac{1}{2}g_{\mu\nu}(\nabla_{\sigma}\Phi)(\nabla^{\sigma}\Phi)-g_{\mu\nu}V(\Phi)\right]\ .
\end{equation}
The equation of motion for the axion field is the modified Klein-Gordon
equation given by 
\begin{equation}
\Box\Phi-\frac{dV(\Phi)}{d\Phi}=-\frac{\alpha}{4}\tilde{R}R\ ,
\end{equation}
where $\Box$ is the d'Alembertian operator defined by $\Box\Phi\equiv\nabla_{a}\nabla^{a}\Phi$.

\section{Setup}

The potential of the axion field is given by 
\begin{equation}
V(\Phi)\equiv\frac{1}{2}m^{2}\Phi^{2}\ ,
\end{equation}
where $m$ is the mass of the axion. First of all, we must solve the
equations of motion in the homogeneous background spacetime 
\begin{eqnarray}
ds^{2} & = & g_{\mu\nu}dx^{\mu}dx^{\nu}\nonumber \\
 & = & a^{2}(\eta)\left(-d\eta^{2}+\delta_{ij}dx^{i}dx^{j}\right)\ .
\end{eqnarray}
The axion depends only on the conformal time $\eta$, that is, $\Phi(x^{\mu})\equiv\Phi(\eta)$.
Thus, we have the modified Klein-Gordon equation 
\begin{equation}
\Phi''+2\frac{a'}{a}\Phi'+a^{2}m^{2}\Phi=0\ ,\label{eq:KGeq}
\end{equation}
where the prime denotes the derivative with respect to the conformal
time $\eta$. These equations can be solved numerically. However,
since the time scale of the expansion of the universe is much longer
than the oscillation time scale for the mass range $m\simeq10^{-14}\sim10^{-10}\,\,{\rm eV}$,
we can ignore the cosmic expansion in the Klein-Gordon equation. Hence,
the scale factor can be put as $a(\eta)=1$. Now, we have the solution
as 
\begin{equation}
\Phi=\Phi_{0}\cos(m\eta)\ ,
\end{equation}
where $\Phi_{0}$ is determined by the energy density of the axion
field as 
\begin{align}
\Phi_{0}= & \,\,\frac{\sqrt{2\rho}}{m}\\
\simeq & \,\,2.1\times10^{7}\,\,{\rm eV}\,\nonumber \\
 & \hspace{1em}\times\left(\frac{10^{-10}\,\,{\rm eV}}{m}\right)\,\sqrt{\frac{\rho}{0.3\,\,{\rm GeV/cm^{3}}}}\,.
\end{align}
Here, we normalized by the energy density of the dark matter in the
halo of the Galaxy. Of course, in general, the axions do not need
to dominate the dark matter component.

Now, we move on to the analysis of equation of motion for GWs. The
GWs can be described by the metric 
\begin{eqnarray}
ds^{2}\simeq-d\eta^{2}+\delta_{ij}dx^{i}dx^{j}+h_{ij}dx^{i}dx^{j}\ ,
\end{eqnarray}
where the spatial tensor $h_{ij}$ is transverse and traceless. We
work in the Fourier space and consider GWs with the wave number vector
$\bm{k}=(k^{1},k^{2},k^{3})$. We can define the unit vector $\bm{n}\equiv\bm{k}/k$
with $k\equiv|\bm{k}|$. Then, the transverse condition is rewritten
as $n^{i}h_{ij}=0$ in the Fourier space. The perturbed gravitational
field can be expressed as 
\begin{equation}
h_{ij}(\eta,\bm{k})=h_{+}(\eta,\bm{k})\,e_{ij}^{+}(\bm{n})+h_{\times}(\eta,\bm{k})\,e_{ij}^{\times}(\bm{n})\ ,
\end{equation}
where $h_{+}(\eta,\bm{k})$ and $h_{\times}(\eta,\bm{k})$ are linear
polarization modes. Here, the polarization tensors, $e_{ij}^{+}(\bm{n})$
and $e_{ij}^{\times}(\bm{n})$, are defined by 
\begin{align}
e_{ij}^{+}(\bm{n}) & \equiv u_{i}u_{j}-v_{i}v_{j}\hspace{1em}\text{and}\hspace{1em}e_{ij}^{\times}(\bm{n})\equiv u_{i}v_{j}+v_{i}u_{j}.
\end{align}
where the two orthogonal unit vectors, $\bm{u}$ and $\bm{v}$, satisfy
the relation, $\bm{n}=\bm{u}\times\bm{v}$. In order to discuss parity
violation, it is useful to define circular polarization tensors 
\begin{align}
e_{ij}^{{\rm R}}(\bm{n}) & =\frac{1}{\sqrt{2}}\left(e_{ij}^{+}(\bm{n})+ie_{ij}^{\times}(\bm{n})\right)\ ,\\
e_{ij}^{{\rm L}}(\bm{n}) & =\frac{1}{\sqrt{2}}\left(e_{ij}^{+}(\bm{n})-ie_{ij}^{\times}(\bm{n})\right)\ .
\end{align}
Then, the perturbed gravitational field $h_{ij}$ is expressed by
\begin{equation}
h_{ij}=h_{{\rm R}}\ e_{ij}^{{\rm R}}(\bm{n})+h_{{\rm L}}\ e_{ij}^{{\rm L}}(\bm{n})\ ,
\end{equation}
where the circular polarization modes are defined by 
\begin{align}
h_{{\rm R}}\equiv\,\,\frac{1}{\sqrt{2}}\left(h_{+}-ih_{\times}\right)\ ,\quad h_{{\rm L}}\equiv\,\,\frac{1}{\sqrt{2}}\left(h_{+}+ih_{\times}\right).
\end{align}
By using these expressions and the relation 
\begin{eqnarray}
i\epsilon_{ilm}n_{l}e_{mj}^{{\rm R/L}}(\bm{n})=\pm e_{ij}^{{\rm R/L}}(\bm{n})\ ,
\end{eqnarray}
the gravitational wave equations can be diagonalized as 
\begin{equation}
{\displaystyle h_{{\rm A}}''+\frac{\epsilon_{{\rm A}}\delta\,\cos(m\eta)}{1+\epsilon_{{\rm A}}\frac{k}{m}\delta\,\sin(m\eta)}k\,h_{{\rm A}}'+k^{2}h_{{\rm A}}}=0\ ,\label{Basic}
\end{equation}
where we defined the dimensionless parameter $\delta$ as 
\begin{equation}
\delta\equiv\,\,\frac{\alpha}{\kappa}m^{2}\Phi_{0}\ .
\end{equation}
The parameter $\delta$ characterize the behavior of the gravitational
wave resonance. The capital latin index represents the each parity
state $A={\rm R}$ or ${\rm L}$ and $\epsilon_{{\rm A}}$ is defined
by 
\begin{equation}
\epsilon_{{\rm A}}\equiv\,\,\begin{cases}
1 & {\rm A}={\rm R}\ ,\\
-1 & {\rm A}={\rm L}\ .
\end{cases}
\end{equation}
In the following, we will show Eq.(\ref{Basic}) exhibits the parametric
resonance.

\section{Resonant amplification of GWs }

Suppose that the gravitational waves from some violent astrophysical
events go through our galaxy to reach us. Thus, GWs are affcted by
the coherent oscillation of the axion dark matter.

\begin{figure}
\center\includegraphics[scale=0.8]{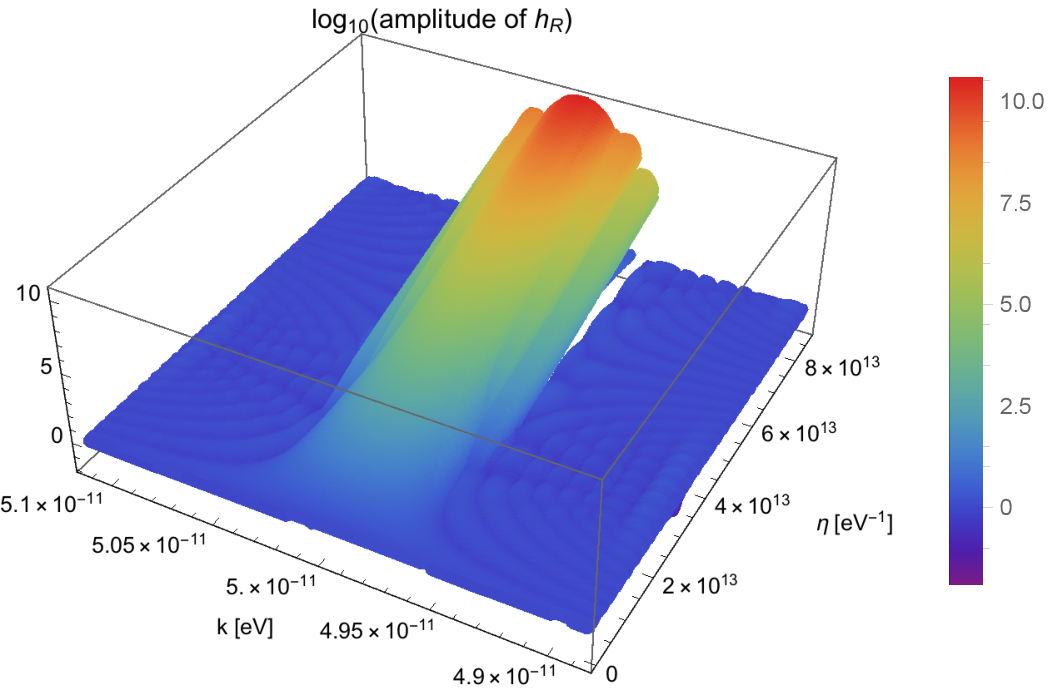}

\medskip{}

\center\includegraphics[scale=0.8]{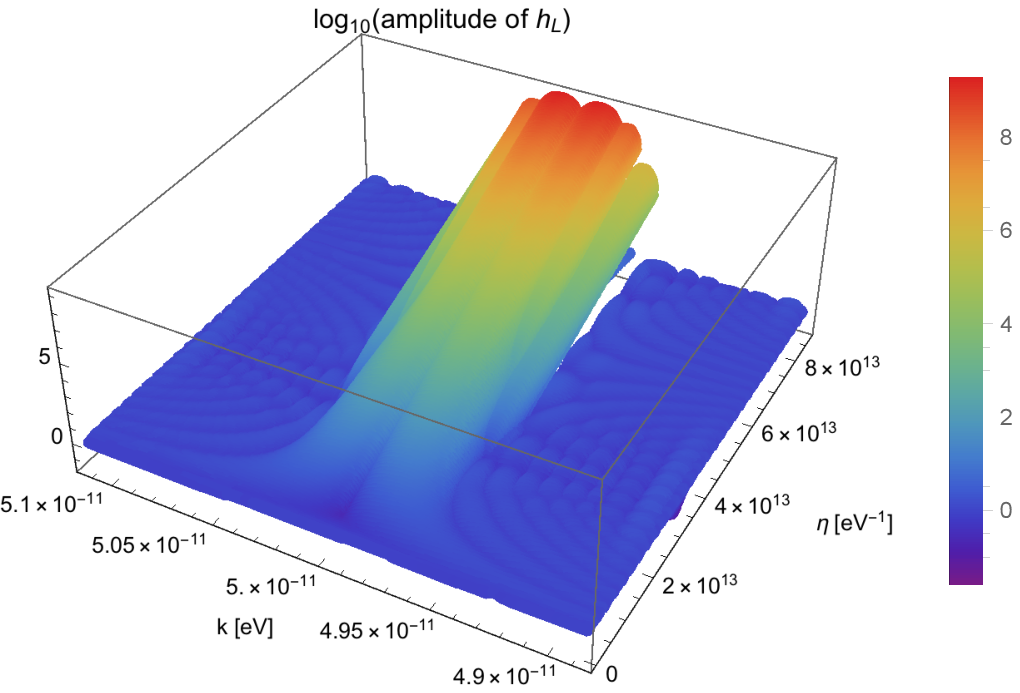}

\caption{The growth of the amplitude of GWs is plotted for $\ell=10^{8}\,{\rm km},\,\,m=10^{-10}\,{\rm eV},\,\,\rho=0.3\times10^{6}\,{\rm GeV/cm^{3}}$. }
\label{fig-:1} 
\end{figure}

\begin{figure}
\center\includegraphics[scale=0.30]{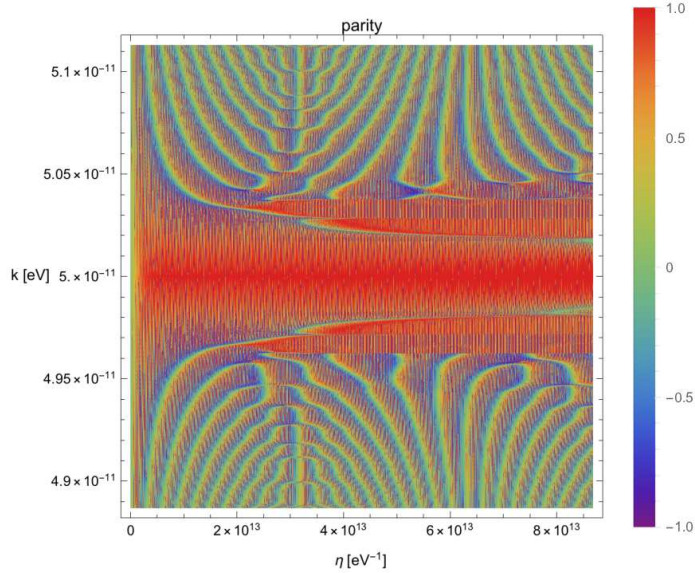}

\caption{The growth of the parity-violation is plotted for $\ell=10^{8}\,{\rm km},\,\,m=10^{-10}\,{\rm eV},\,\,\rho=0.3\times10^{6}\,{\rm GeV/cm^{3}}$. }
\label{fig-:2} 
\end{figure}

We numerically solved Eq.(\ref{Basic}) and plotted the results. In
Fig.\ref{fig-:1}, we plotted the time evolution of the amplitude
of each mode at the resonance frequency. We see the growth rate depends
on the chirality, which stems from the parity violation. In Fig.\ref{fig-:2},
the time evolution of the degree of the circular polarization 
\begin{equation}
{\rm parity}(\eta)\equiv\frac{|h_{{\rm R}}|^{2}-|h_{{\rm L}}|^{2}}{|h_{{\rm R}}|^{2}+|h_{{\rm L}}|^{2}}.
\end{equation}
is plotted. We can see the parity violation clearly from Fig.\ref{fig-:2}.

We can also give analytical estimates. From Eq.(\ref{Basic}), we
find the first resonance wave-number $k_{{\rm r}}$ is given by $k_{{\rm r}}=\,\,{m}/{2}$.
Here, its frequency $f_{{\rm r}}$ can be converted into 
\begin{equation}
f_{{\rm r}}=\frac{k_{{\rm r}}}{2\pi}\simeq\,\,1.2\times10^{4}\,\,{\rm Hz}\,\,\left(\frac{m}{10^{-10}\,\,{\rm eV}}\right).
\end{equation}
Note that this value lies in the detectable range by the ground based
interferometer detectors for GWs. Following the standard analysis
of the parametric resonance, we obtain the width of the resonance
as 
\begin{equation}
\frac{m}{2}-\frac{m}{8}\delta\lesssim k_{{\rm r}}\lesssim\frac{m}{2}+\frac{m}{8}\delta.
\end{equation}
In the case of the black hole binary system, gravitational waves are
superposition of waves with various frequencies. Among them only waves
with the resonant frequency will be enhanced. Hence, the gravitational
waves will have an almost monochromatic frequency. Since the resonance
peak is very sharp, if we detect this signal, we can determine the
mass of the axion dark matter. We can also calculate the growth rate
of GWs due to the resonance by the axion oscillation. The maximum
growth rate $\Gamma_{{\rm max}}$ is given by 
\begin{align}
\Gamma_{{\rm max}} & =\,\,\frac{m}{8}\delta\\
 & \simeq\,\,2.8\times10^{-16}\ {\rm eV}\nonumber \\
 & \hspace{1em}\times\left(\frac{m}{10^{-10}\,\,{\rm eV}}\right)^{2}\,\left(\frac{\ell}{10^{8}\,\,{\rm km}}\right)^{2}\sqrt{\frac{\rho}{0.3\,\,{\rm GeV}/{\rm cm}^{3}}}\ .
\end{align}
Thus, the time $t_{\times10}$ when the amplitude become ten times
bigger is estimated as 
\begin{align}
t_{\times10} & \simeq\,\,8.1\times10^{15}\ {\rm eV^{-1}}\nonumber \\
 & \hspace{1em}\hspace{1em}\times\left(\frac{10^{-10}\,\,{\rm eV}}{m}\right)^{2}\,\left(\frac{10^{8}\,\,{\rm km}}{\ell}\right)^{2}\sqrt{\frac{0.3\,\,{\rm GeV}/{\rm cm}^{3}}{\rho}}.
\end{align}

\begin{figure}
\center\includegraphics[scale=0.67]{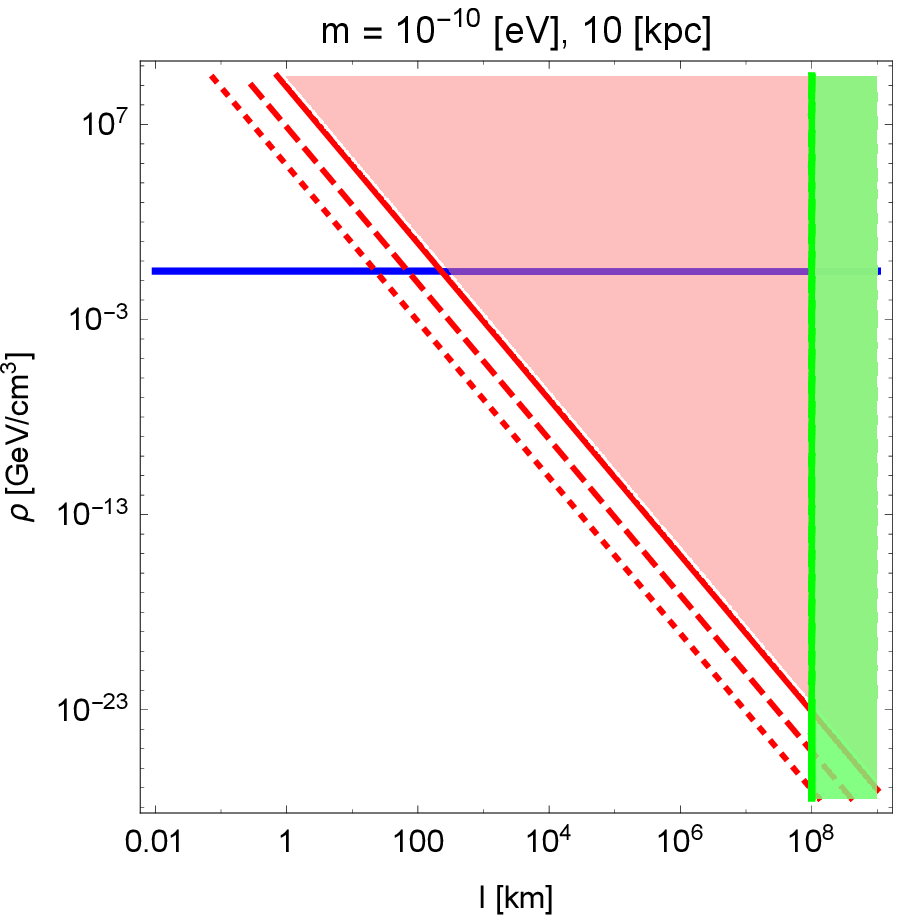}\,\,\,\includegraphics[scale=0.67]{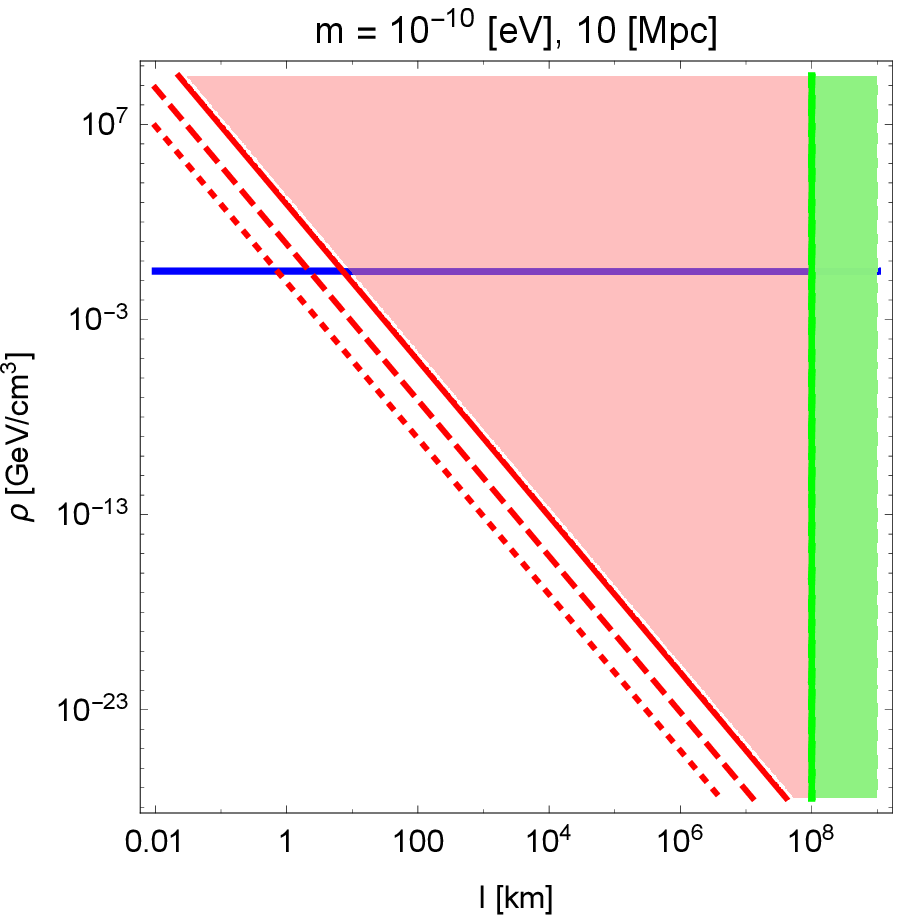}

\label{fig:constraint}\caption{ The constraint on the coupling constant $l$ and the density of the
axion dark matter $\rho$ for $m=10^{-10}\,{\rm eV}$ is shown. The
left one shows the excluded region by the gravitaitonal wave travelling
$10\,{\rm kpc}$ in the axion dark matter, and the right one shows
the excluded region for $10\,{\rm Mpc}$ propagation in the axion
dark matter. The blue line show the local dark matter density $0.3\,{\rm GeV/cm^{3}}$.
The green region is excluded by the observation \cite{AliHaimoud:2011fw}.
The red line represents the ten times enhancement of gravitational
waves. The red dashed line represents the 0.1 times enhancement. The
red dotted line represents the 0.01 times enhancement. Since these
sinatures have never been observed, the upper parameter regions of
these lines are excluded.}
 
\end{figure}

If the distance from the source to the earth is $10\,\,{\rm kpc}$,
the amplitude of GWs can be significantly enhanced. However, there
was no such signal in the past data. This allows us to give constraints
on the coupling $\ell$ and the density $\rho$.  In Fig. \ref{fig:constraint},
we the constraint on the coupling constant $l$ and the density of
the axion dark matter $\rho$. Thus, we obtain the stringent constraint
on the Chern-Simons coupling in the presence of the axion dark mater.

For any axion mass from $10^{-10}\,\,{\rm eV}$ up to $10^{-14}\,\,{\rm eV}$,
we obtain the similar constraint. There are two ways to use the results
we have obtained. If the current upper limit $\ell\sim10^{8}\,\,{\rm km}$
is assumed, the abundance should be constrained. If we assume the
axion is the dark matter, then we obtain the stronger constraint than
the current one $\ell\leq10^{8}\,\,{\rm km}$.

\section{Conclusion}

We studied the string axiverse and parity violation in gravity sector
by considering propagation of GWs in the axion dark matter with the
mass range from $10^{-14}\,\,{\rm eV}$ to $10^{-10}\,\,{\rm eV}$.
It turned out that the axion coherent oscillation induces the parametric
resonance of GWs due to the Chern-Simons term resulting in the circular
polarization of GWs. Thus, the observation of GWs can strongly constrain
the coupling constant of Chern-Simons term and/or the abundance of
the light axions.

In the core of our Galaxy, the dark matter density is higher than
that near the sun by about $10^{3}\sim10^{4}$. Therefore, it is possible
to have more stringent constraint. If the sizable amount of the light
axion exists, the circularly polarized GWs might be observed. If this
happens, it would be a great discovery. If not, as the detection events
increase, the constraints on the Chern-Simons coupling and/or the
abundance of the light axion become tight. In this sense, GWs can
explore the string axiverse and parity violation in gravity.

There are many directions to be studied. We can discuss primordial
gravitational waves in the view of string axiverse. It would be interesting
to extend the present analysis to the gauge Chern-Simons term studied
in \cite{Yoshida:2017fao}. Moreover, the axion-photon conversion
analysis in \cite{Masaki:2017aea} should be extended by taking into
account the axion coherent oscillation. We leave these issues for
future work.

\section*{Acknowledgments}
We would like to thank A. Ito, and R. Kato for useful discussions.
D.Y. was supported by Grant-in-Aid for JSPS Research Fellow and JSPS
KAKENHI Grant Number JP17J00490. J.S. was in part supported by JSPS
KAKENHI Grant Numbers JP17H02894 and JP17K18778, and MEXT KAKENHI
Grant Numbers JP15H05895 and JP17H06359. 

\bibliographystyle{ws-ijmpd}
\bibliography{AxionResonance}

\end{document}